\documentclass[traditabstract]{aa}
\usepackage{graphicx}
\usepackage{natbib}
\usepackage{amssymb}
\usepackage{amsfonts}
\usepackage{amsbsy}
\usepackage{amsmath}

\bibpunct{(}{)}{;}{a}{}{,}

\newcommand{\oneb}{\boldsymbol{1}}
\newcommand{\zerob}{\boldsymbol{0}}
\newcommand{\Ab}{\boldsymbol{A}}
\newcommand{\ab}{\boldsymbol{a}}

\newcommand{\cb}{\boldsymbol{c}}
\newcommand{\Cb}{\boldsymbol{C}}

\newcommand{\Fb}{\boldsymbol{F}}
\newcommand{\nb}{\boldsymbol{n}}
\newcommand{\Ib}{\boldsymbol{I}}

\newcommand{\Hb}{\boldsymbol{H}}

\newcommand{\qb}{\boldsymbol{q}}

\newcommand{\ssb}{\boldsymbol{s}}
\newcommand{\xb}{\boldsymbol{x}}
\newcommand{\rb}{\boldsymbol{r}}

\newcommand{\ub}{\boldsymbol{u}}
\newcommand{\Ub}{\boldsymbol{U}}

\newcommand{\yb}{\boldsymbol{y}}

\newcommand{\wb}{\boldsymbol{w}}
\newcommand{\Wb}{\boldsymbol{W}}

\newcommand{\chib}{\boldsymbol{\chi}}

\newcommand{\sigmab}{\boldsymbol{\sigma}}

\newcommand{\Sigmab}{\boldsymbol{\Sigma}}
\newcommand{\Fmatcb}{\boldsymbol{\mathcal{F}}}

\newcommand{\ph}{\widehat p}

\newcommand{\xh}{\widehat x}
\newcommand{\yh}{\widehat y}
\newcommand{\zh}{\widehat z}

\newcommand{\uph}{\widehat{\upsilon}}

\newcommand{\Abu}{\boldsymbol{\underline{A}}}
\newcommand{\nhu}{\widehat{\underline{n}}}

\newcommand{\rbu}{\boldsymbol{\underline{r}}}
\newcommand{\xhu}{\widehat{\underline{x}}}
\newcommand{\yhu}{\widehat{\underline{y}}}

\newcommand{\xhbu}{\boldsymbol{\widehat{\underline{x}}}}
\newcommand{\yhbu}{\boldsymbol{\widehat{\underline{y}}}}
\newcommand{\chihbu}{\boldsymbol{\widehat{\underline{\chi}}}}
\newcommand{\Fbu}{\boldsymbol{{\underline{F}}}}

\newcommand{\Fmatcbu}{\boldsymbol{{\underline{\mathcal{F}}}}}

\newcommand{\nhb}{\boldsymbol{\widehat n}}
\newcommand{\phb}{\boldsymbol{\widehat p}}
\newcommand{\xhb}{\boldsymbol{\widehat x}}
\newcommand{\yhb}{\boldsymbol{\widehat y}}
\newcommand{\whb}{\boldsymbol{\widehat w}}
\newcommand{\Whb}{\boldsymbol{\widehat W}}
\newcommand{\zhb}{\boldsymbol{\widehat z}}
\newcommand{\uphb}{\boldsymbol{\widehat{\upsilon}}}

\newcommand{\Imatc}{\mathcal{I}}
\newcommand{\Rmatc}{\Re}

\newcommand{\Fmatc}{\mathcal{F}}

\newcommand{\Fmatfb}{\boldsymbol{{\mathfrak F}}}

\begin{document}

   \title{Unevenly-sampled signals: a general formalism of the Lomb-Scargle periodogram}
   \author{R. Vio\inst{1}
    \and
           P. Andreani\inst{2}
    \and
           A. Biggs\inst{3}
          }
   \institute{Chip Computers Consulting s.r.l., Viale Don L.~Sturzo 82,
              S.Liberale di Marcon, 30020 Venice, Italy\\
              \email{robertovio@tin.it},
         \and
		  ESO, Karl Schwarzschild strasse 2, 85748 Garching, Germany\\
                  INAF-Osservatorio Astronomico di Trieste, via Tiepolo 11, 34143 Trieste, Italy\\              
		  \email{pandrean@eso.org}
         \and
		  ESO, Karl Schwarzschild strasse 2, 85748 Garching, Germany\\
                  \email{abiggs@eso.org}
             }

\date{Received .............; accepted ................}

\abstract {The periodogram is a popular tool that tests whether a
  signal consists only of noise or if it also includes other
  components. The main issue of this method is to define a
  critical detection threshold that allows identification of a
  component other than noise, when a peak in the periodogram exceeds
  it. In the case of signals sampled on a regular time grid,
  determination of such a threshold is relatively simple.  When the
  sampling is uneven, however, things are more complicated. The most
  popular solution in this case is to use the {{\it Lomb-Scargle}}
  periodogram, but this method can be used only when the noise is the
  realization of a zero-mean, white (i.e.\ flat-spectrum) random
  process. In this paper, we present a general formalism based on
  matrix algebra, which permits analysis of the statistical
  properties of a periodogram independently of the characteristics of
  noise (e.g.\ colored and/or non-stationary), as well as the
  characteristics of sampling.}

\keywords{Methods: data analysis -- Methods: statistical}
\titlerunning{Periodogram analysis of unevenly-sampled signals}
\authorrunning{R. Vio, P. andreani and A. Biggs}
\maketitle

\section{Introduction}

Spectral analysis is a popular tool for testing whether a given
experimental time series $\{ x(t_0), x(t_1), \ldots, x(t_{M-1}) \}$
contains only noise, i.e.\ $x(t_j) = n(t_j)$, or whether some other
component $s(t)$ is present, i.e.\ $x(t) = s(t) + n(t)$. The classic
approach is to fit the time series with the model function
\begin{equation} \label{eq:model}
x(t_j) = \sum_{k=0}^{N-1} a_k \cos{(2 \pi f_k t_j)} + b_k \sin{(2 \pi f_k t_j}) + n(t_j),
\end{equation}
$j=0, 1, \ldots, M-1$. If, for example, a periodic component $s(t)$ is
present with a frequency $f_l$ close to one in the set $\{ f_k \}$,
then the periodogram $\{ \ph_k \}$,
\begin{equation} \label{eq:period1}
\ph_k = a_k^2 + b_k^2, \qquad k=0, 1, \ldots, N-1, 
\end{equation}
will show a prominent peak close to $k=l$. If $s(t)$ is semi-periodic
or even non-periodic, the situation is more complicated since more
peaks are expected. The main problem with the use of this technique
is the definition of a detection threshold that fixes the contribution
of noise in such a way that, when a peak exceeds it, the presence of a
component $s(t)$ can be claimed. In the case of signals sampled on a
regular time grid, the determination of such a threshold is a
relatively simple procedure, but this is not the case when the
condition of regularity does not apply. In this respect, several
solutions have been proposed \citep[see][and references
  therein]{lom76,fer81, sca82, gil87, ree07, zec09} that, however,
work only under rather restrictive conditions (e.g.\ white and/or
stationary noise) and are difficult to extend to more general
situations.

In this paper, a general formalism is presented that allows
analysis of the statistical properties of periodograms independently
of the specific characteristics of the noise and the sampling of the
signal. In Sec.~\ref{sec:formalization} the formalism is presented
for the case of even sampling and its extension to arbitrary sampling
in Sec.~\ref{sec:generalization}. The usefulness of the proposed
formalism is illustrated in Sec.~\ref{sec:irregular}, where the case
of white noise with a mean different from zero and that of colored
noise is calculated. In Sec.~\ref{sec:fit} the relationship between
the periodogram and the least-squares method is considered. Finally,
on the basis of simulated signals and a real time series, we discuss in
Sec.~\ref{sec:discussion} whether the use of algorithms
specifically developed for computing the periodogram of
unevenly-sampled time series is really advantageous.

\section{Periodogram analysis in the case of even sampling}
\label{sec:formalization}

If a continuous signal $x(t)$ is sampled on a set of $N$ equispaced
time instants $t_0, t_1, \ldots, t_{N-1}$, a time series $x_j$, $j=0,
1, \ldots, N-1$ is obtained. Its discrete Fourier transform (DFT) is
given by
\begin{equation} \label{eq:DFT}
\xh_k = \frac{1}{\sqrt{N}} \sum_{j=0}^{N-1} x_j {\rm e}^{ - i 2 \pi k j / N}, \qquad k=0, 1, \ldots, N-1,
\end{equation}
with $i = \sqrt{-1}$. The sequence $\{ x_j \}$ can be recovered from
$\{ \xh_k \}$ by means of
\begin{equation} \label{eq:IDFT}
x_j = \frac{1}{\sqrt{N}} \sum_{k=0}^{N-1} \xh_k {\rm e}^{ 2 \pi k j/ N}, \qquad j=0, 1, \ldots, N-1. 
\end{equation}
The set $\{ k/N \}_{k=0}^{N-1}$ provides the so-called {\it Fourier
  frequencies}. Implicit in the use of DFT is the assumption that $\{
x_j \}$ is a periodic sequence with period $N \Delta t$ where $\Delta
t = t_{j+1} - t_j$.

In matrix notation, Eqs.~(\ref{eq:DFT}) and (\ref{eq:IDFT}) can be
written in the form
\begin{equation} \label{eq:FDFT}
\xhb = \Fb \xb
\end{equation}
and
\begin{equation} \label{eq:IFDFT}
\xb = \Fb^* \xhb.
\end{equation}
Here, $\xb$ and $\xhb$ are column arrays that contain, respectively,
the sequences $\{ x_j \}$ and $\{ \xh_k \}$, and $\Fb$ is the so-called ``{\it Fourier matrix}'',  which is an $N \times
N$ square, symmetric matrix whose $(k,j)$-entry \footnote{In the following, 
the element in the $n$th row and $m$th column 
of an $N \times M$ matrix $\Ab$ will be indicated with $A_{mn}$ or alternatively with
$(\Ab)_{mn}$, $n=0, 1, \ldots N-1$, $m=0, 1,\ldots, M-1$.} is given by
\begin{equation} \label{eq:F}
F_{k j} = \frac{1}{\sqrt{N}} {\rm e}^{ - i 2 \pi k j /N}.
\end{equation}
The superscript ``${}^*$'' denotes the {\it complex conjugate transpose}.
Matrix $\Fb$ is unitary, i.e.,
\begin{equation} \label{eq:inverse}
\Fb \Fb^* = \Fb^* \Fb = \Ib,
\end{equation}
with $\Ib$ the identity matrix. Another useful property is that
\begin{equation} \label{eq:FF}
\Fb \Fb = \Fb^T \Fb = \Fb \Fb^T = \Fb^T \Fb^T = \Hb 
\end{equation}
with 
\begin{equation} \label{eq:Permut}
\Hb =
\left(
\begin{array}{ccccccc}
1 & 0 & 0 & \cdots & 0 & 0 & 0 \\
0 & 0 & 0 & \cdots & 0 & 1 & 0 \\
0 & 0 & 0 & \cdots & 1 & 0 & 0 \\
\vdots & \vdots & \vdots & \ddots & \vdots & \vdots & \vdots \\
0 & 0  & 1 & \cdots & 0 & 0 & 0  \\
0 & 1 & 0 & \cdots & 0 & 0 & 0 \\
\end{array}
\right).
\end{equation}
By means of Eq.~(\ref{eq:FF}) it can be shown that
\begin{equation} \label{eq:orth}
\Fb_{\Rmatc} \Fb_{\Imatc} = \Fb^T_{\Rmatc} \Fb_{\Imatc} = \Fb_{\Rmatc} \Fb^T_{\Imatc} = \Fb^T_{\Rmatc} \Fb^T_{\Imatc} = \zerob,
\end{equation}
where $\Fb_{\Rmatc} \equiv \Rmatc[\Fb]$ and $\Fb_{\Imatc} \equiv \Imatc[\Fb]$ and where $\Rmatc[.]$ and $\Imatc[.]$ are the real and the 
imaginary parts of a complex quantity. Indeed,
\begin{equation} \label{eq:prop}
\Fb \Fb = \Fb_{\Rmatc} \Fb_{\Rmatc} + \Fb_{\Imatc} \Fb_{\Imatc} + i 2  \Fb_{\Rmatc} \Fb_{\Imatc} = \Hb,
\end{equation}
and $\Hb$ is a real matrix, hence Eq.~(\ref{eq:orth}) has to hold.
From Eq.~(\ref{eq:DFT}) it is also possible to see that $\xh_0$ is a
real quantity. In the case that $N$ is an even number, the same holds
for $\xh_{N_{\dag}+1}$ where $N_{\dag} = \lceil N/2 \rceil$ and
$\lceil z \rceil$ represents the smallest integer greater than
$z$. Finally, dealing with $\xhb$, only half of this array can be
considered, since $\xh_{N-k} = \xh^*_k$, $k=1, 2, \ldots,
N_{\dag}-\delta_{2 N_{\dag}}^N$, where $ \delta_i^j = 1$ if $i=j$ and
zero otherwise. With this notation, the periodogram of $\xb$ is
defined as
\begin{equation} \label{eq:period2}
\ph_k = 2 (\Rmatc[\xhu_k]^2 + \Imatc[\xhu_k]^2) = 2 | \xhu_k |^2, \qquad k=0,1,\ldots, N_{\dag}-1,
\end{equation}
where $\xhu_k$ is the $k$th entry of the array 
$\xhbu^T = [\xh_0, \xh_1, \ldots, \xh_{N_{\dag}-1}]$ \footnote{From now on, if $\rb$ is an $N \times 1$ column array, then
$\rbu$ is a column array that contains the first $N_{\dag} = \lceil N/2 \rceil$ entries of $\rb$, i.e.\ $\rbu = [r_0, r_1, \ldots, r_{N_{\dag}-1}]^T$.
Similarly, if $\Ab$ is an $N \times M$ matrix,
then $\Abu$ is a matrix that contains the first $N_{\dag}$ rows of $\Ab$.}
and $| . |$ denotes the {\it Euclidean} norm.
By means of 
\begin{equation}
\zhb = \left(
\begin{array}{c}
\xhbu_{\Rmatc} \\
\xhbu_{\Imatc}
\end{array}
\right),
\end{equation}
a column array obtained by the column concatenation of
$\xhbu_{\Rmatc} \equiv \Rmatc[\xhbu]$ and $\xhbu_{\Imatc} \equiv \Imatc[\xhbu]$, Eq.~(\ref{eq:period2}) can be rewritten in the form
\begin{equation} \label{eq:period3}
\ph_k = 2 (\zh_k^2 + \zh_{N_{\dag}+k}^2),  \qquad k=0,1,\ldots, N_{\dag}-1.
\end{equation}
In Sect.~\ref{sec:fit} it is shown that this periodogram is equivalent to that obtainable by means the least-squares fit of
model~(\ref{eq:model}) with $M=N$, $t_j = j$ and $f_k = k/N$.

An important point to stress is that, if $\xb$ is the realization of a
(not necessarily Gaussian) random process, then each $\xhu_k$ is given
by the sum of $N$ random variables. This is because of the linearity
of the {\it Fourier operator} $\Fb$.  Thanks to the {\it central limit
  theorem}, therefore, the entries of $\xhb$ can be expected to be Gaussian
random quantities. As a consequence, the entries of $\zhb$ can also be expected to be Gaussian random quantities
with covariance matrix $\Cb_{\zhb}  = {\rm E}[\zhb \zhb^T]$ given by
\begin{equation} \label{eq:Ce}
\Cb_{\zhb} = \left(
\begin{array}{cc}
\Fbu_{\Rmatc} \Cb_{\xb} \Fbu^T_{\Rmatc} & \Fbu_{\Rmatc} \Cb_{\xb} \Fbu^T_{\Imatc} \\
\Fbu_{\Imatc} \Cb_{\xb} \Fbu^T_{\Rmatc} & \Fbu_{\Imatc} \Cb_{\xb} \Fbu^T_{\Imatc}
\end{array}
\right).
\end{equation}
Here, ${\rm E}[.]$ denotes the {\it expectation operator},  $\Cb_{\xb} = {\rm E}[\xb \xb^T]$ is the covariance matrix of
$\xb$, and $\Fbu$
the matrix obtained by the first $N_{\dag}$ rows of the {\it Fourier matrix} $\Fb$.  From
Eqs.~(\ref{eq:Ce}) and (\ref{eq:orth}), it is easy to deduce that, if
$\xb$ is the realization of a standard white-noise process,
i.e., $\Cb_{\xb} = \Ib$, then $\Cb_{\zhb}$ is a diagonal matrix with $(\Cb_{\zhb})_{11}=1$, 
$(\Cb_{\zhb})_{N_{\dag} N_{\dag}}=0$
and $(\Cb_{\zhb})_{kk}=0.5$. In other words, the entries of $\zhb$ are
mutually uncorrelated. In turn, this means that $\Rmatc[\xhu_k]$ is
uncorrelated with $\Imatc[\xhu_k]$. If $\xb$ is a colored (not
necessarily stationary) noise process, i.e.\ $\Cb_{\xb}$ is not a diagonal matrix,
then this holds also for $\Cb_{\zhb}$. However, from
$\xb$ it is possible to obtain an array $\yb$ containing mutually
uncorrelated entries by means of the transformation
\begin{equation} \label{eq:transf}
\yb = \Cb_{\xb}^{-1/2} \xb.
\end{equation}
The matrix $\Cb_{\xb}^{-1/2}$ can be computed via
\begin{equation} \label{eq:eig}
\Cb_{\xb}^{-1/2} = \Ub^T \Sigmab^{-1/2} \Ub
\end{equation}
with $\Ub$ the orthogonal matrix whose columns contain
the eigenvectors of $\Cb_{\yb}$ and $\Sigmab$ a diagonal matrix containing the corresponding eigenvalues $\{ \lambda_l \}_{l=0}^{N - 1}$.
This decomposition is particularly simple and computationally efficient if $\Cb_{\xb}$ is a {\it circulant} matrix since it
can be diagonalized according to
\begin{equation} \label{eq:diag}
\Cb_{\xb} = \Fb^* {\rm diag}[\Fb \cb] \Fb,
\end{equation} 
where ``${\rm diag}[\qb]$'' denotes a diagonal matrix whose diagonal
entries are given by the array $\qb$ and $\cb$ is the first column of
$\Cb_{\xb}$. Because of this, the decomposition~(\ref{eq:eig}) can be
directly computed with $\Ub = \Fb$ and $\Sigmab = {\rm diag}[\cb]$;
hence, $\yb = \Fb^* \Sigmab^{-1/2} \Fb$. This means that the whitening
operation can be performed in the harmonic domain through the
following procedure: a) computation of $\xhb$, i.e.\ the DFT of $\xb$;
b) computation of $\{ \yh_k \} = \{\xh_k / \lambda^{1/2}_k \}$; and c)
the inverse DFT of $\yhb$. A potential difficulty in using
transformation~(\ref{eq:transf}) is that $\Cb_{\xb}$ is required to be
of full rank. Such a condition can be expected to be satisfied in most
practical applications. Some problems can arise if the time step of
the sampling is much shorter than the decorrelation time of
$\nb$ \footnote{The decorrelation time of a random signal $n(t)$ is
  the time interval $\Delta t$ such that two values $n(t_1)$ and
  $n(t_2)$, with $t_2 - t_1 = \Delta t$, can be considered as
  uncorrelated.}. Indeed, some columns (and therefore some rows) of
$\Cb$ could almost be identical i.e., this matrix could become
numerically {\it ill-conditioned}. In this case, the most {\it natural} solution
consists of averaging the data that are close in time.

When the periodogram is used to test whether $\xb = \nb$ vs. $\xb =
\ssb + \nb$, a threshold $L_{\ph_k}$ has to be defined such that, with
a prefixed probability, a peak in $\ph_k$ that exceeds $L_{\ph_k}$
can be expected to not arise because of the noise. This requires 
knowledge of the statistical properties of $\phb$ under the hypothesis
that $\xb = \nb$. The simplest situation is when $\nb$ is the
realization of a standard white-noise process. In fact, since the
entries of $\zhb$ are uncorrelated random Gaussian quantities, from
Eq.~(\ref{eq:period3}) it can be derived that the entries of $\phb$
are (asymptotically) independent quantities distributed according to a $\chi^2_2$
distribution\footnote{$\chi^2_2$ denotes the chi-square distribution
  with two degrees of freedom.}. As a consequence, independent of
the frequency $k$, a threshold $L_{{\rm Fa}}$ can be determined that
corresponds to the level that a peak due to the noise would exceed
with a prefixed probability $\alpha$ when a number $N_f$ of ({\it statistically
  independent}) frequencies are inspected. More specifically, $L_{{\rm
    Fa}}$ is the highest value for which
$1-[1-\exp{(-L_{\ph_k})}]^{N_f} \le \alpha$ \citep{sca82}, in formula
\begin{equation} \label{eq:false}
L_{{\rm Fa}} = \underset{ L_{\ph_k} }{\sup} \left\{ 1-[1-\exp{(-L_{\ph_k})}]^{N_f} \le \alpha \right\}.
\end{equation}
If the entire periodogram is inspected, then $N_f = N_\dag$. Commonly, $L_{{\rm
    Fa}}$ is called the {\it level of false alarm}.

Threshold~(\ref{eq:false}) is not applicable when the noise is
colored. However, the requirement to fix a different level for each
frequency can be avoided if the original signal $\xb$ is transformed
into
\begin{equation} \label{eq:whitey}
\yb = \Cb^{-1/2}_{\nb} \xb.
\end{equation}
Indeed, under the hypothesis that $\xb=\nb$, the entries of $\yb$ are
uncorrelated and unit-variance random quantities. As seen above,
this operation should not be difficult. Some problems emerge if the 
decomposition~(\ref{eq:eig}) is carried out by means of the efficient DFT approach described above
(a necessary approach in the case of very long sequences of data). This is because, in
forming $\Cb_{\nb}$, it is necessary to take the
periodicity of the sequence $\xb$ forced by the DFT into account. In other words,
it is necessary to impose a spurious correlation among the first and
the last entries in $\nb$. For instance, for a stationary noise process,
$\Cb_{\nb}$ is a {\it Toeplitz} matrix, but it has to be approximated
with a {\it circulant} one. When dealing with sampled signals, this is
an unavoidable problem that no technique can completely solve. A
classical solution to relieving this situation is the {\it windowing
  method}, i.e., the substitution of $x_j$ in Eq.~(\ref{eq:DFT}) with
$\eta_j x_j$, where $\{ \eta_j \}$ is some prefixed discrete function
(window) that makes the signal gently reduce to zero at the extreme of
the sampling interval \citep[e.g. see][]{opp89}. This method can be
expected to work satisfactorily only when the decorrelation time of
noise is shorter than the length of the signal.

\section{Periodogram analysis in the case of uneven sampling} \label{sec:generalization}

If a signal $x(t)$ is sampled on an uneven set of time instants, some
problems emerge: it is no longer possible to define a set of "{\it
  natural}" frequencies such as those obtained by the {\it Fourier}
transform. In turn, this implies some ambiguities in the definition of
the Nyquist frequency that, loosely speaking, corresponds to the
highest frequencies that contain information on the signal of interest
\citep[e.g. see][]{vio00, koe06}. As a consequence,
Eq.~(\ref{eq:DFT}) has to be modified. In the following, with no loss
of generality, it is assumed that $\xb(t)$ is sampled at $M$
arbitrary time instants $t_0, t_1, \ldots, t_{M-1}$ with $t_0 = 0$,
$t_{M-1} = M-1$ and the remaining $t_j$ arbitrarily distributed within
this interval. Moreover, a set of $N$ frequencies $k =0, 1, \ldots,
N-1$ is considered with $N \gtreqless M$. Such a set corresponds
to the frequencies that are typically inspected when looking for a
periodicity. However, others can be chosen. With these conditions, a
transformation corresponding to the one given by Eq.~(\ref{eq:FDFT}) is
\begin{equation} \label{eq:irr4}
\xhb = \Fmatcb \xb,
\end{equation}
where
\begin{equation} \label{eq:irr3}
\Fmatc_{k j} = \frac{1}{\sqrt{M}} {\rm e}^{ - i 2 \pi k \tilde{t}_j / N},
\end{equation}
$\tilde{t}_j = t_j/ \Delta_m t$, $\Delta_m t = \gamma \min{ [
    \{t_{j+1} - t_j\}]}$ and $\gamma$ is a real positive number. Because of the normalization 
by $\Delta_m t$, time $\tilde{t}_j$ is expressed in units of the
shortest sampling time interval. Apart from the substitution of the {\it Fourier} matrix $\Fb$
with $\Fmatcb$, the uneven sampling of signals does not modify the formalism introduced in the previous section.
In particular, the covariance matrix $\Cb_{\zhb}$ defined in Eq.~(\ref{eq:Ce}) becomes
\begin{equation} \label{eq:Cei}
\Cb_{\zhb} = \left(
\begin{array}{cc}
\Fmatcbu_{\Rmatc} \Cb_{\xb} \Fmatcbu^T_{\Rmatc} & \Fmatcbu_{\Rmatc} \Cb_{\xb} \Fmatcbu^T_{\Imatc} \\
\Fmatcbu_{\Imatc} \Cb_{\xb} \Fmatcbu^T_{\Rmatc} & \Fmatcbu_{\Imatc} \Cb_{\xb} \Fmatcbu^T_{\Imatc}
\end{array}
\right).
\end{equation}

The $N \times M$ {\it Fourier matrix} $\Fmatcb$ does not have the
properties~(\ref{eq:inverse})-(\ref{eq:prop}). As consequence, and
also in the case that $\xb$ is the realization of a standard
white-noise process i.e.\ $\Cb_{\xb} = \Ib$, matrix $\Cb_{\zhb}$,
\begin{equation} \label{eq:CI1}
\Cb_{\zhb} = \left(
\begin{array}{cc}
\Fmatcbu_{\Rmatc} \Fmatcbu^T_{\Rmatc} & \Fmatcbu_{\Rmatc} \Fmatcbu^T_{\Imatc} \\
\Fmatcbu_{\Imatc} \Fmatcbu^T_{\Rmatc} & \Fmatcbu_{\Imatc} \Fmatcbu^T_{\Imatc}
\end{array}
\right),
\end{equation}
is not diagonal. In general, $\Cb_{\zhb}$ is not even diagonalizable.
For example, this happens when the periodogram of a time series containing $M$ data
is computed on $N$ frequencies with $N > M$ (a typical situation in practical applications).
This implies that
the entries of $\zhb$ cannot be made mutually uncorrelated. Obviously,
the same holds for the entries of $\phb$ as given by
Eq.~(\ref{eq:period3}). As a consequence, although a number $N$ of frequencies are considered in $\phb$, at most only $M/2$
of them are statistically independent\footnote{This is because,
if $N > M$, the rank of the $N \times N$ matrix $\Cb_{\zhb}$ is smaller than or equal to $M$. 
This implies that the array ${\zhb}$ has at most $M$ degrees of freedom.
Since each entry of $\phb$ is given by the sum
of two entries of ${\zhb}$, then a periodogram has at most $M/2$ degrees of freedom.}. 
Particularly troublesome is that, for a given
frequency $k$, $\Rmatc[\xhu_k]$, and $\Imatc[\xhu_k]$ (i.e.\ $\zh_k$
and $\zh_{N_{\dag}+k}$) are also correlated. This makes it difficult
to fix the statistical characteristics of $\ph_k$. In this respect,
two choices are possible. The first consists in the determination, for
each frequency, of the PDF of $\ph_k$. Actually, this is a rather involved 
approach, because $\zh_k$ and $\zh_{N_{\dag} + k}$ have
variance $(C_{\zhb})_{kk}$ and $(C_{\zhb})_{N_{\dag} + k, N_{\dag} +
  k}$, respectively, and covariance $(C_{\zhb})_{k, N_{\dag} +
  k}$. Therefore, once $\zh_k$ and $\zh_{N_{\dag} + k}$ are normalized to unit variance,
each $\ph_k$ is given by the sum of two correlated $\chi_1^2$ random
quantities. Although available in analytical form \citep{sim06}, the resulting PDF is rather complex
and hence difficult to handle \citep[for an alternative approach, see][]{ree07}.
Moreover, there is the additional problem that
$L_{\ph_k}$ changes with $k$. A simpler alternative is the use of two
uncorrelated and unit-variance random quantities, $\uph_k$ and
$\uph_{N_{\dag} + k}$, obtained through the transformation
\begin{equation} \label{eq:norm}
\uphb = \Sigmab^{-1/2}_\star \Ub_\star \zhb_\star. 
\end{equation}
Here, $\uphb^T = [\uph_k, \uph_{N_{\dag} + k}]$, $\zhb_\star^T =
[\zh_k, \zh_{N_{\dag} + k}]$, $\Sigmab_{\star}^{-1/2}$ is a diagonal matrix
whose entries are given by the reciprocal of the square root of
the non-zero eigenvalues of the covariance matrix
\begin{equation} \label{eq:step3}
\Cb_\star = \left(
\begin{array}{cc}
(C_{\zhb})_{kk} & (C_{\zhb})_{k, N_{\dag} + k} \\
(C_{\zhb})_{k, N_{\dag} + k}  & (C_{\zhb})_{N_{\dag} + k, N_{\dag} + k} 
\end{array}
\right),
\end{equation}
zero otherwise,
and $\Ub_\star$ is an orthogonal matrix that contains the
corresponding eigenvectors\footnote{The diagonal elements $\lambda_1$
  and $\lambda_2$ of $\Sigma_*$ can be trivially computed through the
  solution of the quadratic equation $\lambda^2 - {\rm tr}[\Cb_\star]
  + {\rm det}[\Cb_\star] = 0$, with ${\rm tr[.]}$ and ${\rm det}[.]$
  denoting the {\it trace} and {\it determinant} operators. The arrays
  $\ub_1$ and $\ub_2$, which constitute the columns of $\Ub_*$, can be
  obtained by solving the equations $(\Cb_\star - \lambda_l \Ib) \ub_l
  = \zerob$, $l=1,2$.}. Indeed, if the periodogram is defined as
\begin{equation} \label{eq:step1}
\ph_k = \uph_k^2 + \uph^2_{N_{\dag} + k}, \qquad k=0,1, \ldots, N_{\dag} -1,
\end{equation}
then each $\ph_k$ is given by the sum of two independent,
unit-variance, Gaussian random quantities. As a consequence, the
corresponding PDF is, independently of $k$, a $\chi^2_2$ whose {\it
  cumulative distribution function} (CDF) is the exponential
function. This permits determining the statistical significance of $\ph_k$ 
for a {\it specified} frequency $k$. Things become more
complex if $N_f$ frequencies are
inspected when looking for a peak. Indeed, also after the operation~(\ref{eq:norm}), it
happens that ${\rm E}[\ph_k \ph_l] \neq 0$ for $k \neq l$, i.e., the
frequencies of the periodogram remain mutually correlated. This is an
unavoidable problem. Because of it, $N_f$ does not correspond to the
number of {\it independent frequencies}, so the {\it level of
  false alarm}~(\ref{eq:false}) cannot be computed. However,
since $L_{\ph_k}$ is the same for all the frequencies, an
upper limit can be fixed for $L_{{\rm Fa}}$ by setting $N_f = \lceil
M/2 \rceil$. The periodogram obtained by means of Eq.~(\ref{eq:step1})
corresponds to the original {\it Lomb-Scargle} periodogram.

Since the transformation~(\ref{eq:transf}) does not depend on the
characteristics of the signal sampling, the strategy of following in the
case that $\xb$ is the realization of (not necessarily stationary)
colored noise is simply the one in
Sec.~\ref{sec:formalization} i.e.\ transformation of $\xb$ to an
array $\yb$ with uncorrelated entries. After that, the {\it
  Lomb-Scargle} periodogram can be computed. It is worth noticing that
this simple result has been possible thanks to a formulation of the
problem in the time domain and the use of the matrix notation. The
same results could have been obtained by following the popular
approach of working in the harmonic domain but at the price of a much
more difficult derivation.

\section{Two examples} \label{sec:irregular}

To illustrate the usefulness and the simplicity of the proposed
formalism in handling different situations from the classical ones, we
show two examples in this section.

The first consists of a periodogram of a mean-subtracted time
series. The evaluation of the reliability of a peak in the periodogram
of a signal $\xb$ requires that (under the {\it null} hypothesis $\xb = \nb$) $\nb$ be the realization of a
zero-mean noise process. In most experimental situations, this
condition is not fulfilled and one works with a centered
(i.e.\ mean-subtracted) version $\chib$ of $\xb$. This, however,
introduces some (often neglected) problems. The case where $\xb$ is
the realization of a discrete white noise process with variance
$\sigma^2_{\xb}$ has been considered several times in the literature.
An example is the paper by \citet{zec09} where a rather elaborate
solution is presented. With the approach proposed here, a simpler
solution can be obtained if one takes into consideration that the
subtraction of the mean from $\xb$ forces a spurious correlation among
the entries of $\chib$ in such a way that the covariance matrix $\Cb_{\chib}=E[ \chib \chib^T ]$ is given by
\begin{equation} \label{eq:chib}
\Cb_{\chib} = \sigma^2_{\xb} \left( \Ib - \frac{\large{\oneb}}{M} \right),
\end{equation}
where $M$ is number of entries of $\xb$ and $\large{\oneb}$ an $M \times M$ matrix with every entry equal to unity. 
Since this matrix is singular, it cannot be diagonalized and therefore $\chib$ cannot be whitened. 
In any case, if in Eq.~(\ref{eq:Cei}) matrix $C_{\xb}$ is substituted for $\Cb_{\chib}$ and one sets
\begin{equation}
\zhb = \left(
\begin{array}{c}
\chihbu_{\Rmatc} \\
\chihbu_{\Imatc}
\end{array}
\right),
\end{equation}
then it is a trivial matter to decorrelate
$\zh_k$ and $\zh_{N_{\dag} + k}$ by means of
Eqs.~(\ref{eq:norm})-(\ref{eq:step3}) and to compute the
periodogram through Eq.~(\ref{eq:step1}). This result can be easily
extend to the case where, because of measurement errors, each entry of
$\xb$ has its own variance $\sigma^2_{x_j}$ and a weighted mean is
subtracted from the data sequence i.e.\ $\chi_j = x_j - \sum _l
\eta_j x_l / \sum_l \eta_l$, with $\eta_l = 1/
\sigma^2_{x_l}$. Indeed, it is sufficient to substitute $\Cb_{\chib}$
as given by Eq.~(\ref{eq:chib}) with
\begin{equation}
\Cb_{\chib} = {\rm diag}[\sigmab^2] - \frac{\large{\oneb}}{\sum_l \eta_l},
\end{equation}
where $\sigmab^2=[\sigma^2_{x_0}, \sigma^2_{x_1}, \ldots, \sigma^2_{x_{N-1}}]^T$.
The rest of the procedure remains unmodified.

\begin{figure*}
        \resizebox{\hsize}{!}{\includegraphics{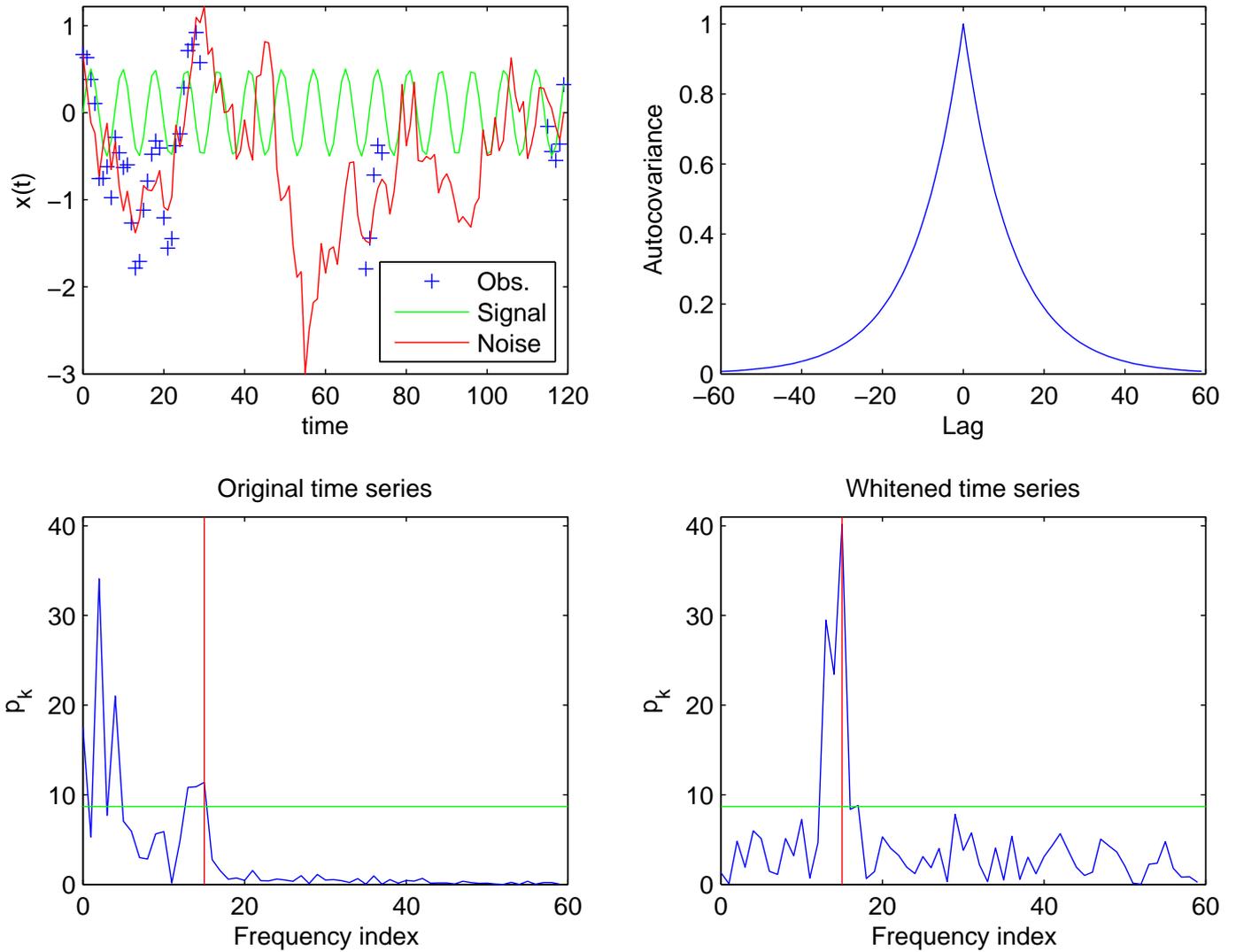}}
        \caption{Results concerning the numerical experiment, presented in Sec.~\ref{sec:irregular},
        on the detection of a sinusoidal component in colored noise. The top left panel shows
	  an observed time series (blue crosses) obtained through the simulation of
        signal $\xb = \{ x_j \}_{j=0}^{119}$ with $x_j = 0.5 \sin{(2 \pi f j)} + n_j$, $f=0.127$,
        on a regular grid of $120$ time instants but with the data in the ranges $[31~70]$ and $[76~115]$.
        Here, $\nb$ (red line) is the realization of a discrete,
        zero-mean, colored noise process whose autocovariance function is given in the top right panel. For comparison, the sinusoidal component 
        is also plotted (green line). The bottom left panel shows the 
        {\it Lomb-Scargle} periodogram of the original sequence $\xb$ 
        computed on $120$ frequencies $k=0,1/120, \ldots, 119/120$, whereas the bottom right 
        panel shows the {\it Lomb-Scargle} periodogram corresponding to its whitened version. In both cases, only the first 
        $60$ frequencies are shown and the vertical red line corresponds to the frequency of the sinusoidal component. The horizontal green line
        in the bottom panels provides the threshold corresponding to a $0.01$ {\it level of false alarm} (number of independent frequencies $N_f=60$),
        i.e.\ the probability that the periodogram
        of a pure noise signal exceeds such a threshold by chance is $1\%$.}
        \label{fig:color}
\end{figure*}

The second example consists of zero-mean colored noise. The
improvement in the quality of the results obtainable with the approach
presented in the previous section is visible in
Fig.~\ref{fig:color}. The top left panel shows a discrete signal $x_j
= 0.5 \sin{(2 \pi f j)} + n_j$, $f=0.127$, simulated on a regular grid
of $120$ time instants but with missing data in the ranges $[31~70]$
and $[76~115]$. Here, $\nb$ is the realization of a discrete,
zero-mean, colored noise process whose autocovariance function is
given in the top right panel. From the bottom left panel, it is
evident that {\it Lomb-Scargle} periodogram of the original sequence $\xb$
provides rather ambiguous results concerning the presence of a
sinusoidal component. On the other hand, such component is well visible
in the bottom right panel that shows the {\it Lomb-Scargle} periodogram of the sequence
$\yb = \Cb_{\nb}^{-1/2} \xb$.

\section{Periodogram and least-squares fit of sinusoids} \label{sec:fit}

The formalism proposed here is also useful in the context of more
theoretical questions (but with important practical implications).
For example, a point often overlooked in the astronomical literature is
the relationship between the periodogram and the least-squares fit of
sine functions. Often these two methods are believed to be
equivalent. Actually, this is true only when the sampling is regular
and the frequencies of the sinusoids are given by the {\it Fourier}
ones. Indeed, if $t_j = \tilde{t}_j$ and $f_k = k / N$, $k=0, 1,
\ldots, N-1$, then Eq.~(\ref{eq:model}) can be written in the
form
\begin{equation} \label{eq:ls}
\xb -  \Fmatfb^T \ab = \nb,
\end{equation} 
with
\begin{equation}
\Fmatfb = \frac{2}{\sqrt{N}} \left(
\begin{array}{c}
\Fmatcb_{\Rmatc} \\
\Fmatcb_{\Imatc}
\end{array}
\right),
\end{equation}
and $\ab = [a_0, a_1, \ldots, a_{N-1}, b_0, b_1, \ldots, b_{N-1}]^T$. The least-squares solution $\bar{\ab}$ of system~(\ref{eq:ls}) is given by
\begin{equation} \label{eq:lsfit}
\bar{\ab} = (\Fmatfb \Fmatfb^T)^+ \Fmatfb \xb,
\end{equation}
where ``${}^+$'' denotes {\it Moore-Penrose pseudo-inverse} \citep{bjo96}. In the case of even sampling,
i.e.\ when $\Fmatcb_{\Rmatc} = \Fb_{\Rmatc}$ and $\Fmatcb_{\Imatc} = \Fb_{\Imatc}$, it happens that
\begin{equation}
\bar{\ab} = \Fmatfb \xb.
\end{equation}
In other words, coefficients $\{ a_k \}$ and $\{ b_k \}$, as given by
the least-squares approach, can be obtained through the DFT of $\xb$,
because, as shown by means of
Eqs.~(\ref{eq:orth}), $(\Fmatfb \Fmatfb^T)^+ \Fmatfb = \Fmatfb$. In
the case of uneven sampling, this identity is not fulfilled. Any kind
of periodogram computed through Eq.~(\ref{eq:irr4}) and the
least-squares fit of sine functions has to be expected to give
different results. Moreover, as only under the
two above-mentioned conditions do the sine functions constitute an
orthonormal basis, the least-squares fit of a single sine function per
time does not in general provide the same result as the simultaneous
fit of all the sinusoids as in Eq.~(\ref{eq:lsfit})
\citep[e.g. see][page 450]{ham73}. In particular, if an
unevenly-sampled signal is given by the contribution of two or more
sinusoids, the one-at-a-time fit of a single sine function provides
biased results. This also holds for the {\it Lomb-Scargle}
periodogram, which is equivalent to the least-squares fit of a
single sinusoid with a specified frequency, with the constraint that
the corresponding coefficients ``$a$'' and ``$b$'' are uncorrelated
\citep{sca82, zec09}. 

\section{Discussion} \label{sec:discussion}

As demonstrated in Sec.~\ref{sec:generalization}, when noise has
arbitrary statistical characteristics, the computation of the
periodogram of an unevenly-sampled signal requires two steps:
\begin{itemize}
\item Whitening and standardization of the noise component (in this way a signal $\yb$ is obtained);
\item Computation of the {\it Lomb-Scargle} periodogram of $\yb$. 
\end{itemize}
The first step, unavoidable even in the case of regular sampling, is a
computationally expensive operation. Therefore, for time series
containing more than a few thousand data points, dedicated
algorithms exploiting the specific structure of $\Cb_{\nb}$
(e.g.\ often this matrix is of banded type) have to be developed for
implementing Eq.~(\ref{eq:whitey}). However, this problem is
beyond the aim of the present paper. The second step is much less
time consuming. Indeed, in the case of time series containing some
thousands of points and when the periodogram has to be computed on a
similar number of frequencies, the direct implementation of
Eqs.~(\ref{eq:irr4})-(\ref{eq:step1}) results in a few seconds of
computation time only. In other words, in many practical situations,
no dedicated algorithm is really necessary. However, fast algorithms have been proposed 
for very long time series \citep{pre92}.

The last issue that has to be considered is to which extent the use of
the {\it Lomb-Scargle} periodogram is really advantageous. Indeed, the
action of the algorithms dealing with uneven sampling is essentially
directed, for each frequency $k$, to force $\ph_k$ to be the sum of
two independent Gaussian random quantities. However, although not
clearly emphasized, it has already been pointed out
that this operation is not critical \citep[e.g. see][]{sca82}. It can
be expected that a periodogram computed simply through
\begin{equation} \label{eq:pow}
\ph_k = 2 | \yhu_k |^2,
\end{equation}
with $\yhbu = \Fmatcbu \yb$, is often very close to the one given by
Eq.~(\ref{eq:step1}). A rigorous demonstration of this fact is
difficult because of its strict dependence on the specific
sampling. However, with the help of some numerical experiments and of
the formalism introduced here, some insights are possible. In
particular, we consider the covariance matrix $\Cb_{\zhb}$ of a
white noise signal when sampled on different uneven time grids.

\subsection{Numerical simulations}

\begin{figure*}
        \resizebox{\hsize}{!}{\includegraphics{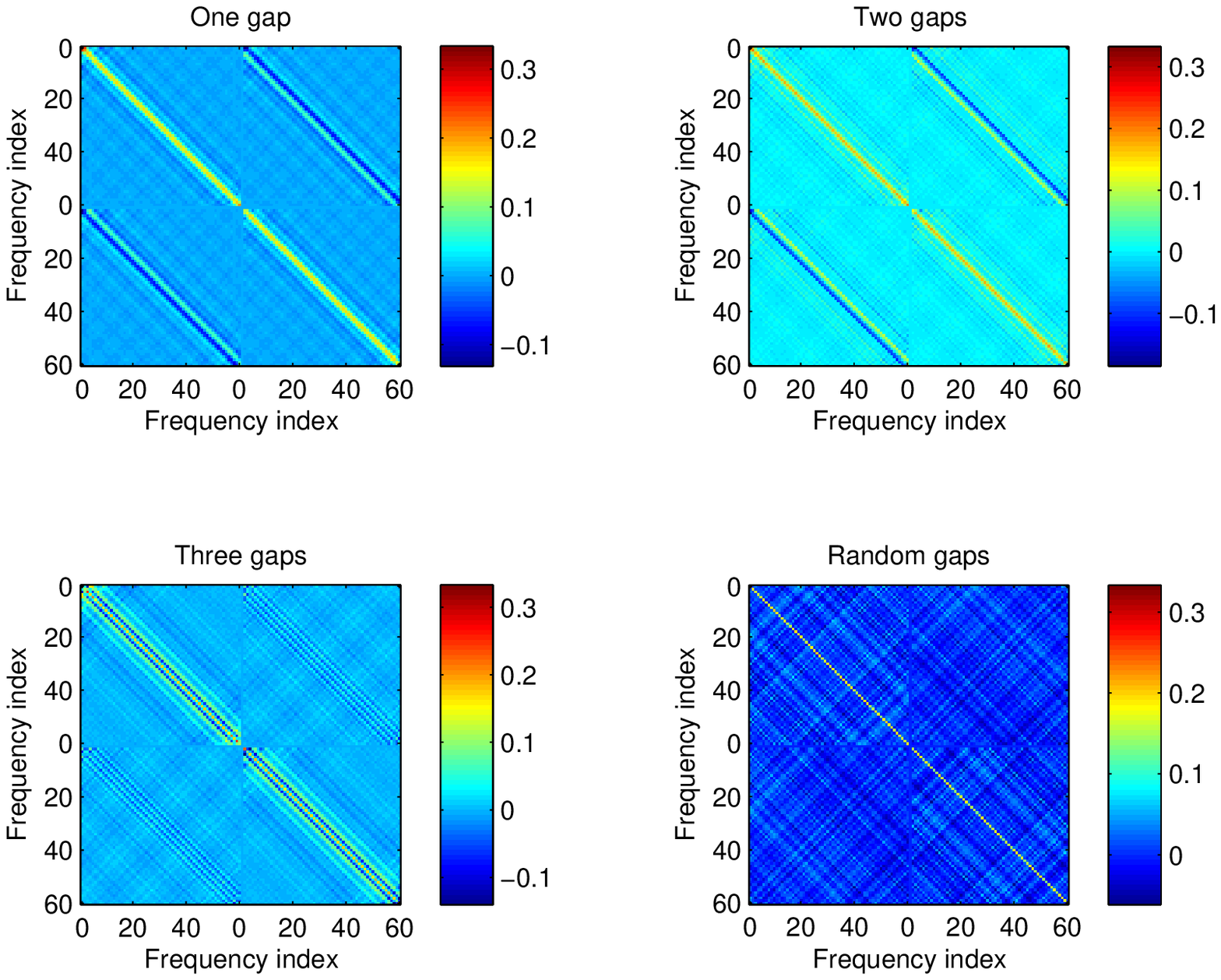}}
        \caption{Covariance matrix of the real and the imaginary parts, say $\Rmatc[\xhu_k]$ and $\Imatc[\xhu_k]$ of the Fourier transform 
        $\{ \xh_k \}$ of an unevenly sampled white-noise signal $\xb$ (i.e.\ matrix $\Cb_{\zhb}$ as given by Eq.~(\ref{eq:CI1})). 
        In particular, $\xb$ is assumed to be the realization of a
        zero-mean, unit-variance, white-noise process sampled at $120$ time instants regularly spaced but with $80$ missing data. Three different 
        cases have been considered with missing data in the following ranges: a) $[31~110]$ (top left panel), b) $[31~70]$, and $[76~115]$
        (top right panel), c) $[6~25]$, $[36~75]$ and $[96~115]$ (bottom left panel). In a fourth case, the missing data are randomly distributed
        (bottom right panel). Because of the small size of the figure, it is necessary to stress that for each panel, none of the prominent 
        diagonal structures visible in the top right, as well in the bottom left quadrant, correspond to the main diagonal of the quadrant itself
        (i.e.\ none of them provide the covariance between $\Rmatc[\xhu_k]$ and $\Imatc[\xhu_k]$ that is seen in Fig.~\ref{fig:covd}).}
        \label{fig:cov}
\end{figure*}

\begin{figure*}
        \resizebox{\hsize}{!}{\includegraphics{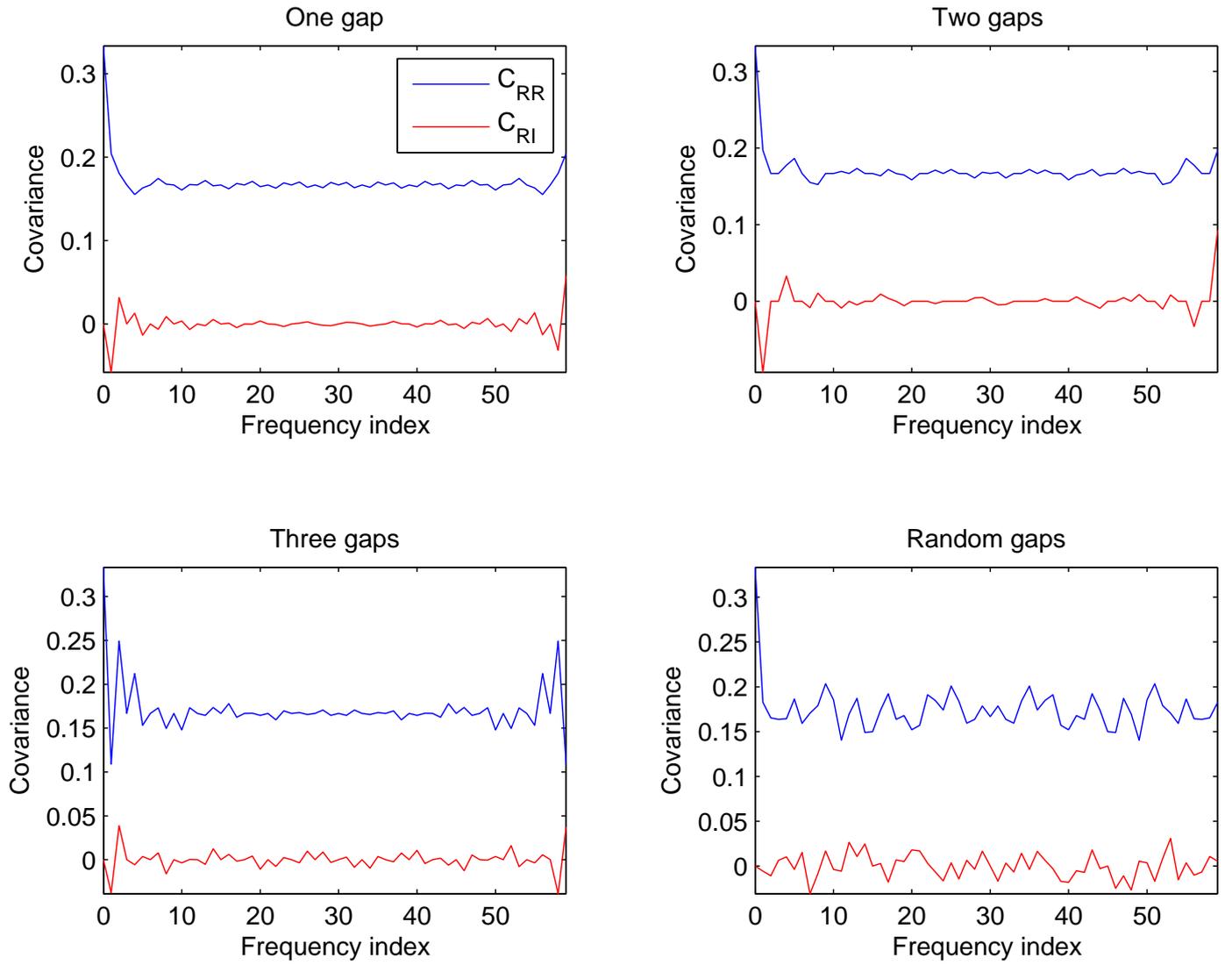}}
        \caption{Variance of $\Rmatc[\xhu_k]$ and covariance of $\Rmatc[\xhu_k]$ with $\Imatc[\xhu_k]$ as a function of the frequency $k$ 
        for the experiments in Fig.~\ref{fig:cov}. The blue line corresponds to the main diagonal of the top left quadrant in each panel 
        of Fig.~\ref{fig:cov}, whereas the red one corresponds to the main diagonal in the top right, as well in the bottom left quadrant.}
        \label{fig:covd}
\end{figure*}

In our simulations, we take the realization of a zero-mean,
unit-variance, Gaussian white-noise process $\nb$ sampled on $M_s=120$ time
regularly-spaced instants, but with $80$ missing data (i.e.\ $M=40$). The available
signal $\xb$ can be written in the form 
\begin{align}
\xb &= \Wb \nb, \label{eq:window1}\\
\Wb & = {\rm diag}[\wb], \label{eq:window2}
\end{align}
where $\Wb = {\rm diag}[\wb]$ and $\wb$ is an array whose entries are equal to one in
correspondence to a value of $\nb$ that is available and zero
otherwise. In this case, Eq.~(\ref{eq:DFT}) can be written as
\begin{align} 
\xhb & = \Fb \Wb \nb \label{eq:irr1} \\
& = \Fmatcb \nb,
\end{align}
with $\Fmatcb = \Fb \Wb$. Three different cases have been considered
where the missing data have time indices in the ranges a) $[31~110]$,
b) $[31~70]$ and $[76~115]$, c) $[6~25]$, $[36~75]$ and $[96~115]$,
whereas they are randomly distributed in a fourth case . The related
covariance matrices $\Cb_{\zhb}$, computed through Eq.~(\ref{eq:CI1})
with $N=M_s$, are shown in Fig.~\ref{fig:cov}, whereas
Fig.~\ref{fig:covd} displays the corresponding main diagonal of the
blocks $\Fmatcbu_{\Rmatc} \Fmatcbu^T_{\Rmatc}$ and $\Fmatcbu_{\Rmatc}
\Fmatcbu^T_{\Imatc}$. Especially from Fig.~\ref{fig:covd} it is
evident that, for an arbitrary frequency $k$, the covariance between
$\Rmatc[\xhu_k]$ and $\Imatc[\xhu_k]$ is quite close to zero. This
means that each entry of $\phb$, as given by Eq.~(\ref{eq:pow}), can
be assumed to be distributed according to a $\chi^2_2$.  From
Fig.~\ref{fig:cov} it is also evident that, in the case of gaps
present in the sampling pattern, there are nearby frequencies $k$ and
$l$ for which not only $\Rmatc[\xhu_k]$ and $\Rmatc[\xhu_l]$ are 
mutually correlated, but also $\Rmatc[\xhu_k]$ with $\Imatc[\xhu_l]$ and
$\Rmatc[\xhu_l]$ with $\Imatc[\xhu_k]$. These correlations,
especially those between the real and the imaginary components,
disappear in the case of random sampling.

\subsection{Interpretation of the results of the simulations}

\begin{figure*}
        \resizebox{\hsize}{!}{\includegraphics{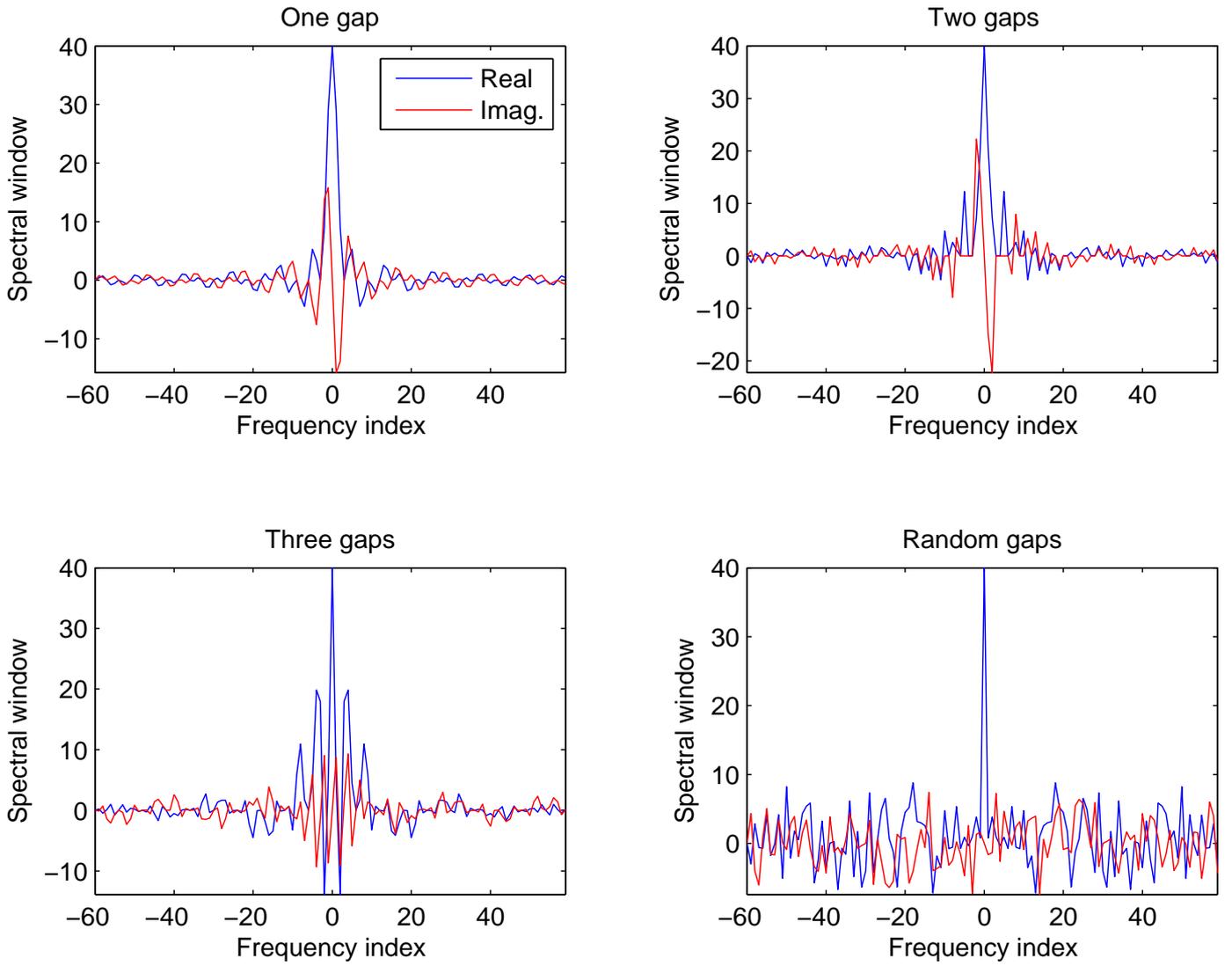}}
        \caption{Real (blue line) and imaginary (red line) parts of the spectral windows $\whb$ that have been used to simulate 
        the irregular sampling of the signal 
        in the experiments corresponding to Fig.~\ref{fig:cov} (i.e., $\whb$ is the Fourier transform of the sampling pattern $\wb$ of $\xb$,
        see Eqs.~(\ref{eq:window1})-(\ref{eq:window2})).}
        \label{fig:windows}
\end{figure*}

\begin{figure*}
        \resizebox{\hsize}{!}{\includegraphics{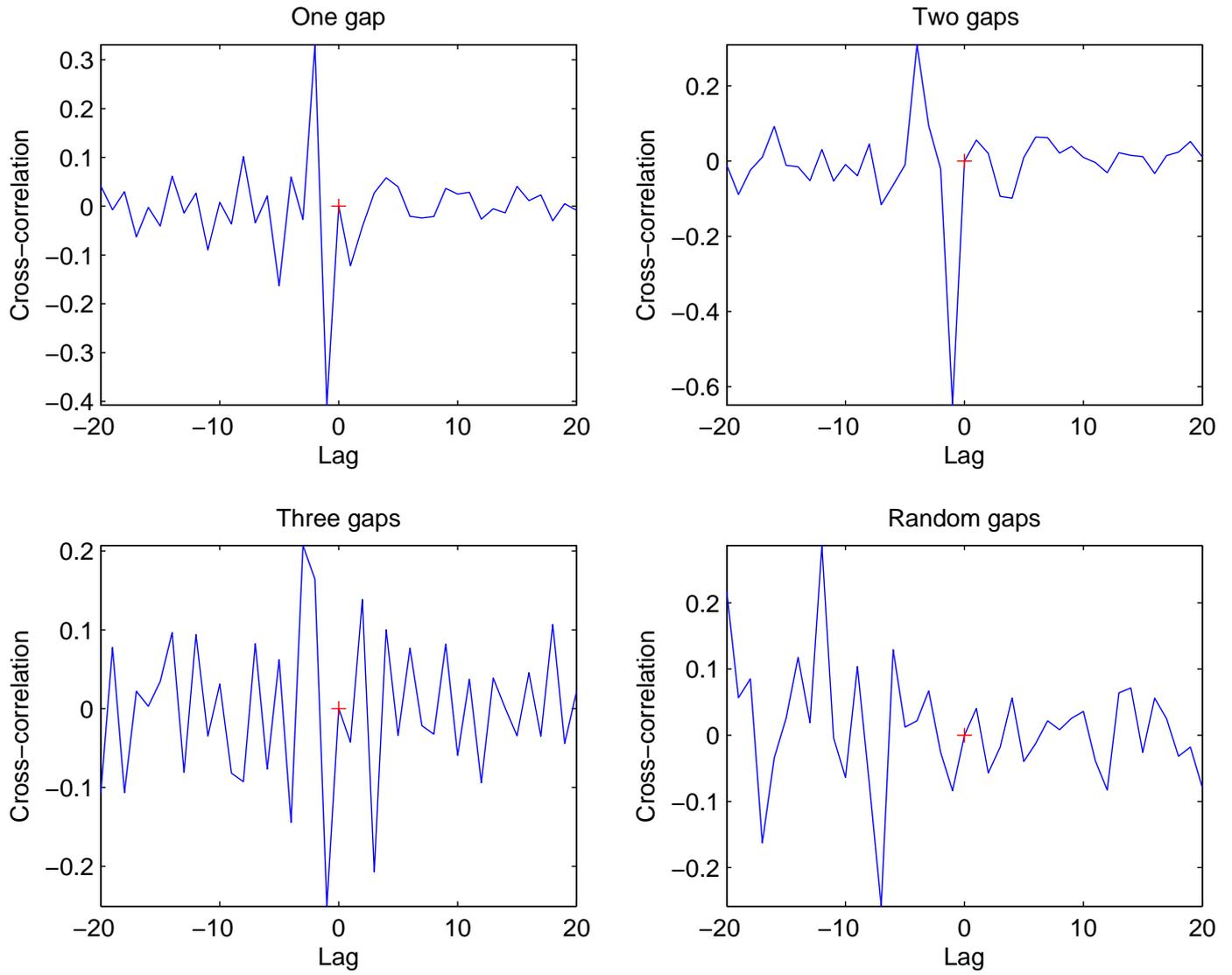}}
        \caption{Cross-correlation between the real and the imaginary parts of the spectral windows $\whb$ that are shown in Fig.~\ref{fig:windows}.
        The red cross in each panel corresponds to the point $(0,0)$.}
        \label{fig:cross}
\end{figure*}

\begin{figure*}
        \resizebox{\hsize}{!}{\includegraphics{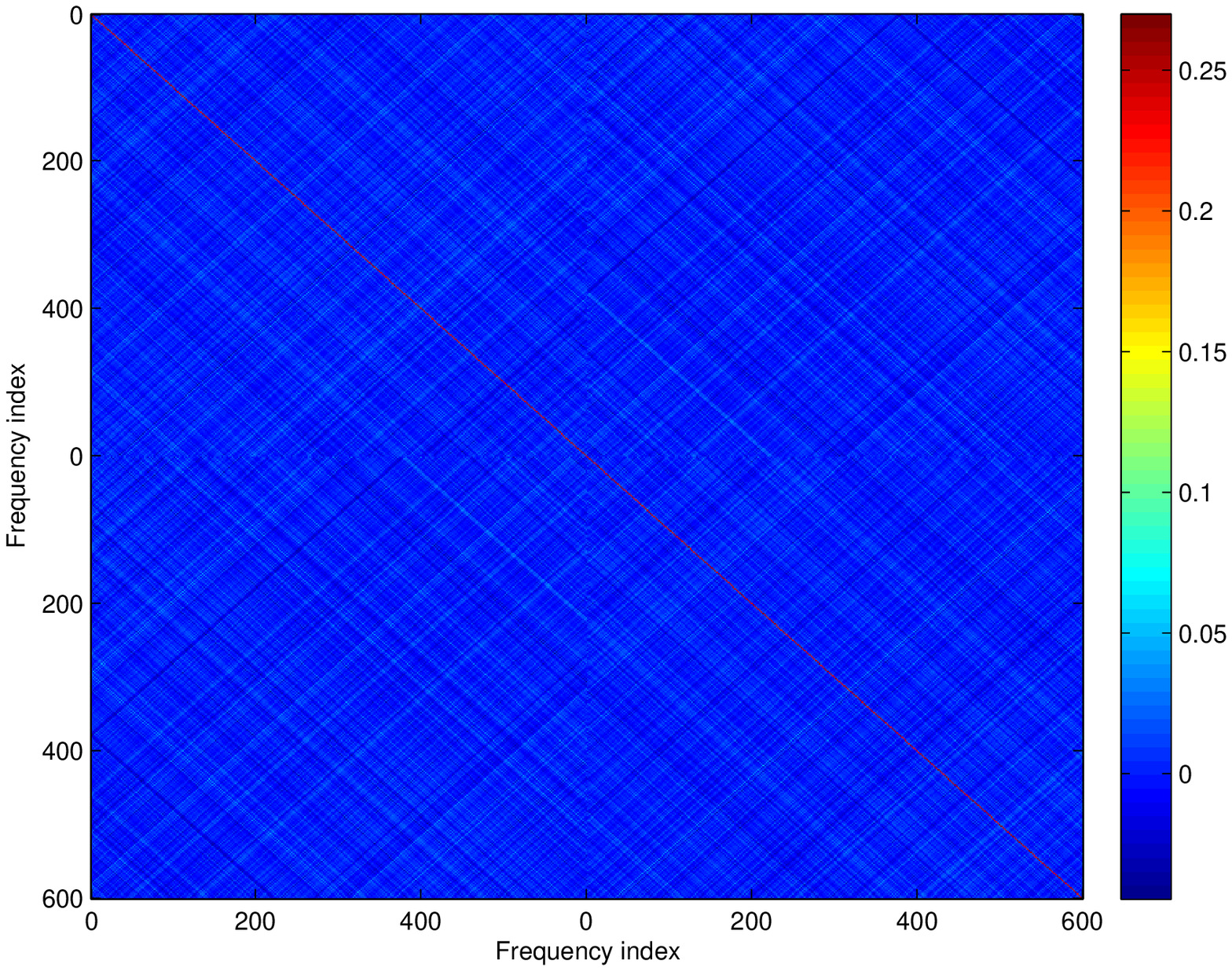}}
         \caption{Covariance matrix of the real and the imaginary parts, say $\Rmatc[\xhu_k]$ and $\Imatc[\xhu_k]$ of the Fourier transform 
        $\{ \xh_k \}$ of an unevenly sampled white-noise signal $\xb$ with periodic gaps (i.e.\ matrix $\Cb_{\zhb}$ as given by Eq.~(\ref{eq:CI1})).
        In particular, $\xb$ is the realization of a discrete zero-mean, unit-variance, white-noise process sampled
        on a grid of $600$ time instants randomly distributed  (i.e.\ not rebinned) according to the cyclic sampling pattern shown in the bottom panel
        of Fig.~\ref{fig:covp}. In total, 1200 frequencies have been considered.}
        \label{fig:covpr}
\end{figure*}

\begin{figure*}
        \resizebox{\hsize}{!}{\includegraphics{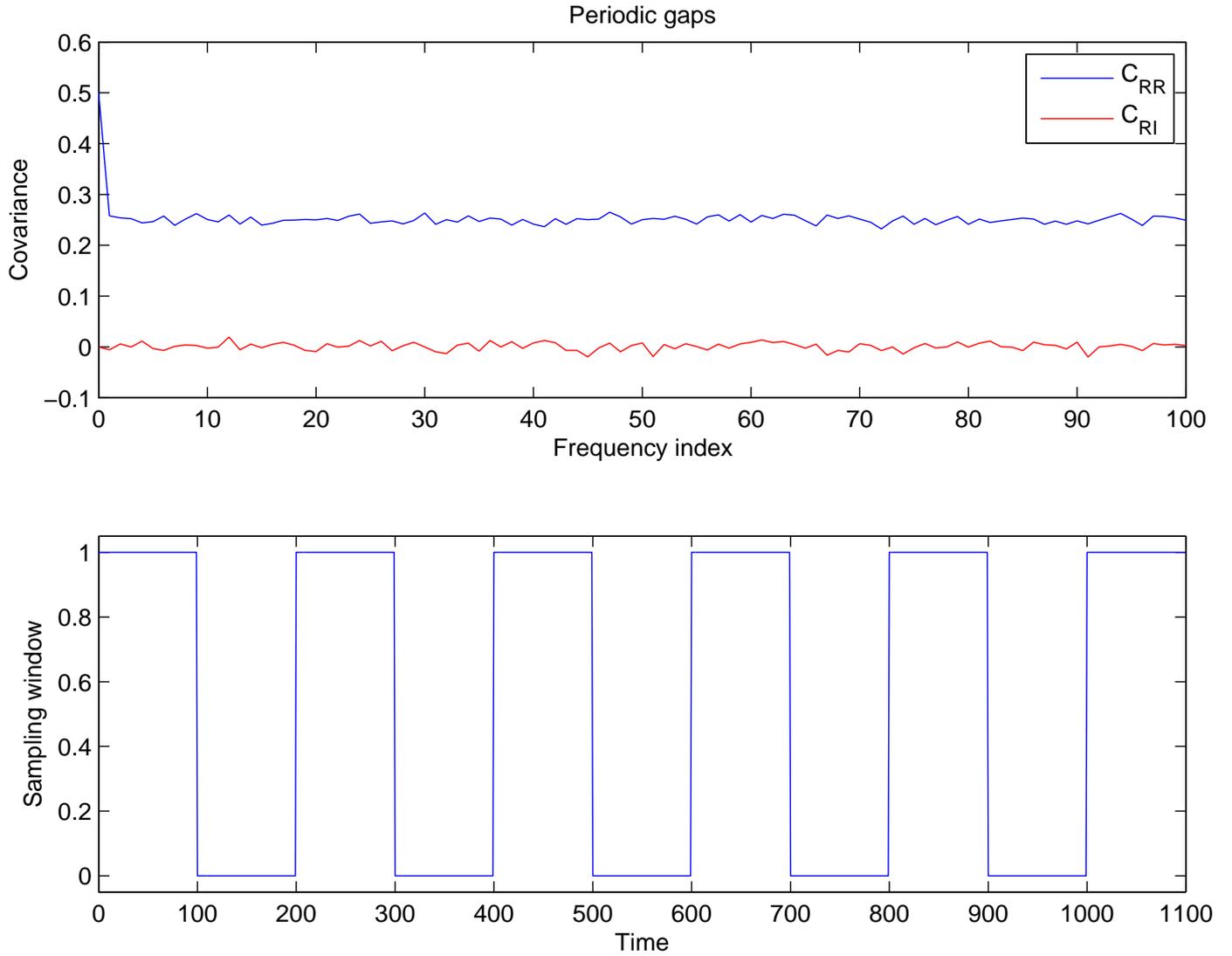}}
        \caption{Top panel: variance of $\Rmatc[\xhu_k]$ and covariance of $\Rmatc[\xhu_k]$ with $\Imatc[\xhu_k]$ (only the first $100$
        frequencies are shown) for the experiment 
        concerning the realization of a discrete zero-mean, unit-variance, white-noise process on $600$ time instants randomly distributed 
        (i.e.\ not rebinned) according to a cyclic sampling. In total, $1200$ frequencies have been considered (see text in Sec.~\ref{sec:discussion}).
        Bottom panel: cyclic sampling used in the experiment.}
        \label{fig:covp}
\end{figure*}

To understand these results, it is necessary to take into
account that Eq.~(\ref{eq:irr1}) can be rewritten in the form
\begin{align} 
\xhb & = \Fb \Wb \Fb^* \Fb \nb, \\
& = \Whb \nhb.
\end{align}
As $\Wb$ is a (singular) diagonal matrix, $\Whb$ is a (singular)
circulant matrix. This implies that $\xhb$ is given by the {\it
  circular convolution} of $\nhb$ (the DFT of the original signal,
inclusive of the missing data) with the {\it spectral window} $\whb$
(the DFT of the sampling pattern). Since for two arbitrary
frequencies, say $k$ and $l$, $\Rmatc[\nhu_k]$, $\Imatc[\nhu_k]$,
$\Rmatc[\nhu_l]$, and $\Imatc[\nhu_l]$ are mutually independent, any
correlation existing between $\Rmatc[\xhu_k]$, $\Imatc[\xhu_k]$,
$\Rmatc[\xhu_l]$, and $\Imatc[\xhu_l]$ is induced by the
correlation between the entries of $\Rmatc[\whb]$ and
$\Imatc[\whb]$. Now, as shown in Fig.~\ref{fig:windows}, in the
presence of gaps the real and the imaginary parts of $\whb$ are both
only significantly different from zero in a narrow interval of
frequencies surrounding the origin, i.e., corresponding to the lowest
frequencies ($\wb$ can be interpreted as low-pass filter). This
produces the correlations observed among $\Rmatc[\xhu_k]$,
$\Rmatc[\xhu_l]$, $\Rmatc[\xhu_l]$ and $\Imatc[\xhu_k]$ for nearby
frequencies $k$ and $l$. On the other hand, in the case of random
sampling $\whb$ mimics the behaviour of white noise, making the
correlations less important. Moreover, Fig.~\ref{fig:cross}
shows that, in the case of sampling with gaps, the cross-correlation
between $\Rmatc[\whb]$ with $\Imatc[\whb]$ is significant but not
centered at zero lag. This explains why the quantities
$\Rmatc[\xhu_k]$ and $\Imatc[\xhu_l]$, $k \neq l$ can be strongly
correlated in spite of the small correlation between
$\Rmatc[\xhu_k]$ and $\Imatc[\xhu_k]$. 
Similar arguments also hold for the case of
sampling with periodic gaps (rather common in astronomical experiments). 
Indeed, in practical applications the period of these gaps is somewhat  smaller than the mean sampling time step
of an uninterrupted sequence of data. As a consequence, both
$\Rmatc[\whb]$ and $\Imatc[\whb]$ present a sharp and narrow peak in correspondence to
a frequency close to the origin. Actually, the {\it spectral window} of a sampling with periodic
gaps is also characterized by the presence of {\it aliases}. However, these aliases too are
sharp and narrow, and their importance decreases for increasing frequencies. 
The combination of these facts leads again to
the quantities $\Rmatc[\xhu_k]$, $\Rmatc[\xhu_l]$, $\Rmatc[\xhu_l]$, and $\Imatc[\xhu_k]$
almost being uncorrelated if
the frequencies $k$ and $l$ are not sufficiently close enough.
Moreover, since with periodic gaps the cross-correlation
between $\Rmatc[\whb]$ with $\Imatc[\whb]$ can also be significant, but not
centered on zero lag, then for each frequency $k$ 
the correlation between $\Rmatc[\xhu_k]$ and $\Imatc[\xhu_k]$ is negligible.
These considerations are confirmed by Figs.~\ref{fig:covpr} and \ref{fig:covp}.
Figure~\ref{fig:covpr} displays the covariance matrix $\Cb_{\zhb}$, computed through Eq.~(\ref{eq:CI1}),
of a discrete zero-mean, unit-variance, and white-noise process sampled on $600$
time instants randomly distributed (i.e.\ not rebinned) along six
cycles (i.e.\ $100$ points per cycle) of the sampling pattern shown in
the bottom panel of Fig.~\ref{fig:covp}; $1200$ frequencies have been
considered. The top panel of Fig.~\ref{fig:covp} displays the
main diagonal of the blocks $\Fmatcbu_{\Rmatc} \Fmatcbu^T_{\Rmatc}$
and $\Fmatcbu_{\Rmatc} \Fmatcbu^T_{\Imatc}$.

\begin{figure*}
        \resizebox{\hsize}{!}{\includegraphics{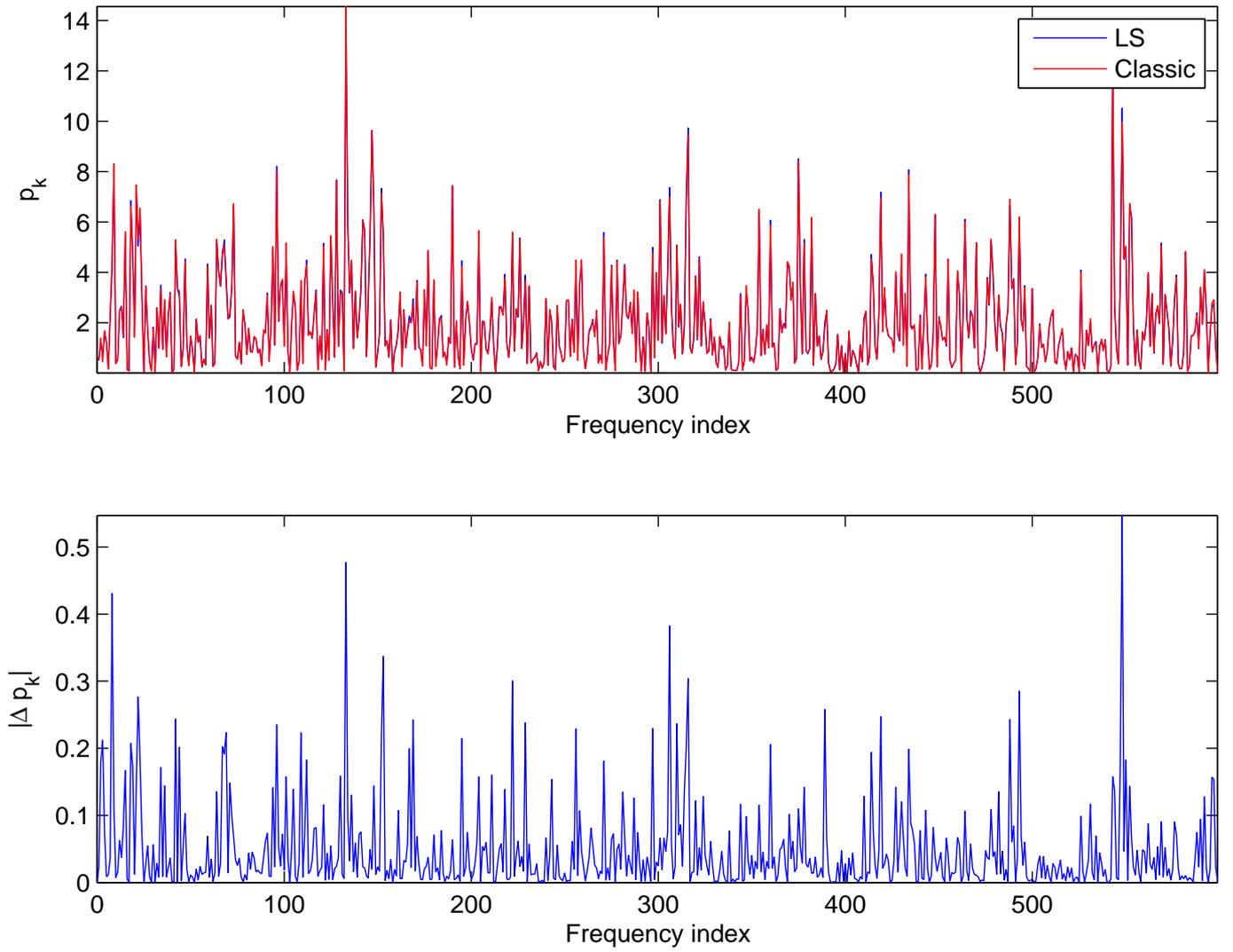}}
        \caption{The top panel shows the {\it Lomb-Scargle} periodogram and the classic periodogram given by $\ph_k = 2 |\xhu_k| ^2$  
        for the experiment in Fig.~\ref{fig:covp}. The bottom panel shows the smallness of their absolute difference.}
        \label{fig:example}
\end{figure*}

The numerical simulations indicate that the consequences of
unevenly-sampled data seem to concern the number of independent
frequencies in $\phb$ rather than the correlation between $\Rmatc[\xhu_k]$
and its imaginary counterpart $\Imatc[\xhu_k]$. At present, no
general method has been developed to deal with this
problem. However, it has been pointed out in the literature that the
number of independent frequencies is not a critical parameter to test
the significance level of a peak in $\phb$.
In particular, empirical arguments indicate that this number can be
safely set to $M/2$ \citep[e.g. see][]{pre92}.  The conclusion is that forcing each entry
of $\phb$ to be the sum of two independent Gaussian quantities only has
minor effects. This is shown by Fig.~\ref{fig:example}
where the top panel shows that the
{\it Lomb-Scargle} periodogram of the time series with periodic gaps 
considered above 
is quite similar to that provided simply by
$\ph_k = 2 |\xhu_k|^2$ with $\xhbu = \Fmatcbu \xb$. This is
evident in the bottom panel of the same figure where the similarity of
the two periodiograms is demonstrated by their absolute difference.

\begin{figure*}
        \resizebox{\hsize}{!}{\includegraphics{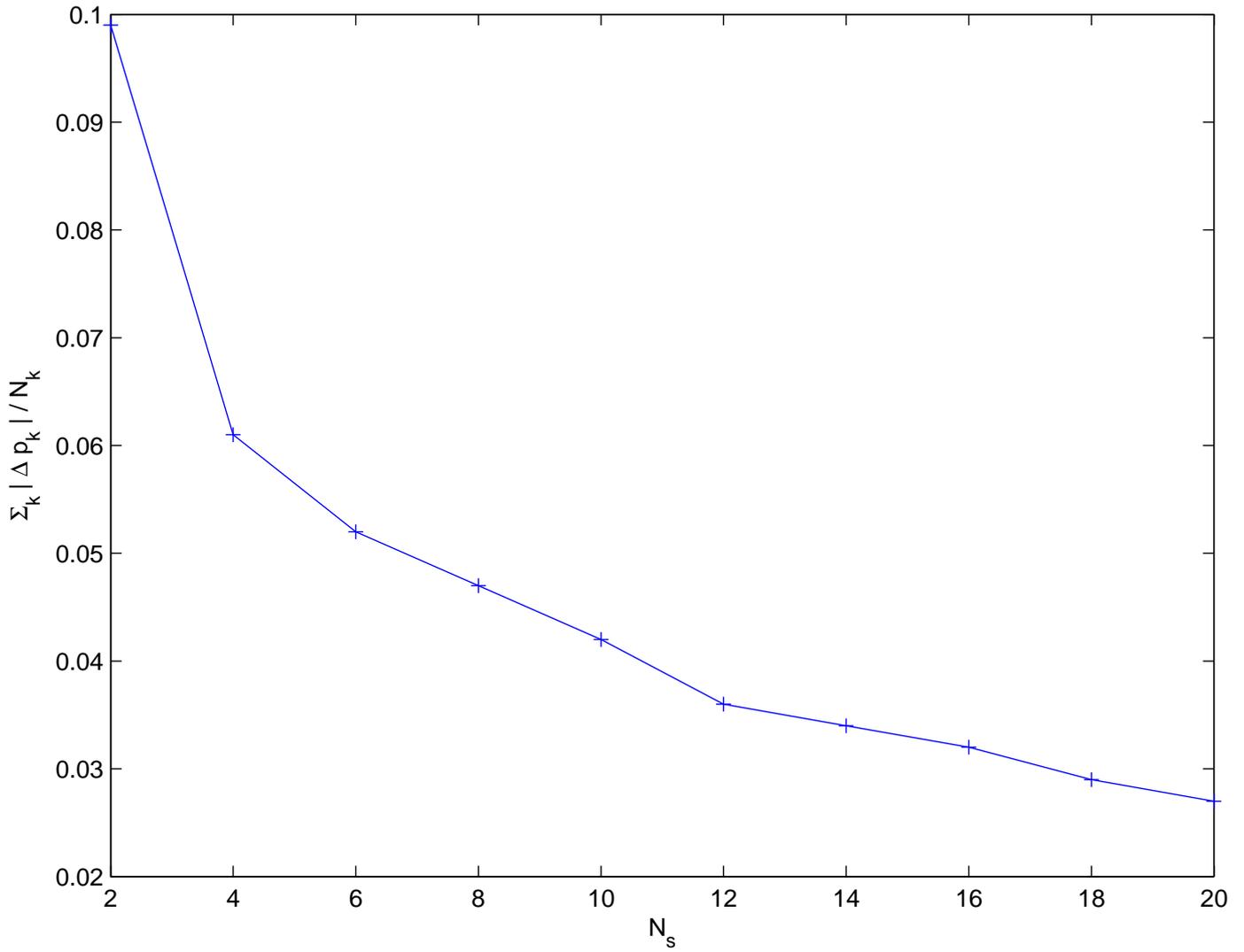}}
        \caption{Mean absolute difference $\sum_k | \Delta \ph_k | / N_k$ ($N_k = $ number of frequencies)
        between the {\it Lomb-Scargle} periodiogram and the classic periodogram given by $\ph_k = 2 |\xhu_k| ^2$ as a function of the number $N_s$ of
        cyclic sampling patterns (Fig.~\ref{fig:example} shows the case with $N_s = 6$).}
        \label{fig:long}
\end{figure*}

A final point to underline, which has important practical
implications, is that for long time series the small differences
visible in Fig.~\ref{fig:example} should decrease. Indeed, as seen
above, $\ph_k$ will be significantly correlated with $\ph_l$ only if
the two frequencies $k$ and $l$ are close enough. The only exception
is represented by the frequencies at the extremes of the frequency
domain where the assumption of periodic signal intrinsic to DFT forces
a spurious correlation. For longer and longer time series, this
spurious correlation will affect a smaller and smaller fraction of
frequencies and, as consequence, a larger and larger fraction of them
will be mutually independent. This is shown in Fig.~\ref{fig:long}
where, in the context of the previous experiment, the mean absolute
difference between the two periodograms is plotted as a function of $N_s$, 
the number of cyclic sampling
patterns ($N_s = 6$ in Fig.~\ref{fig:example}). This argument also
explains why in many practical situations the number of independent frequencies 
can be safely fixed to $M/2$. In conclusion, only in
the case of signals that contain a small number of data, the {\it Lomb-Scargle} 
periodogram can be expected to exhibit noticeable differences from the 
periodogram given by
Eq.~(\ref{eq:pow}). Often, using it does not change anything. Comparable
results can be expected with less sophisticated approaches.

\subsection{Application to an astronomical time series}

\begin{figure*}
        \resizebox{\hsize}{!}{\includegraphics{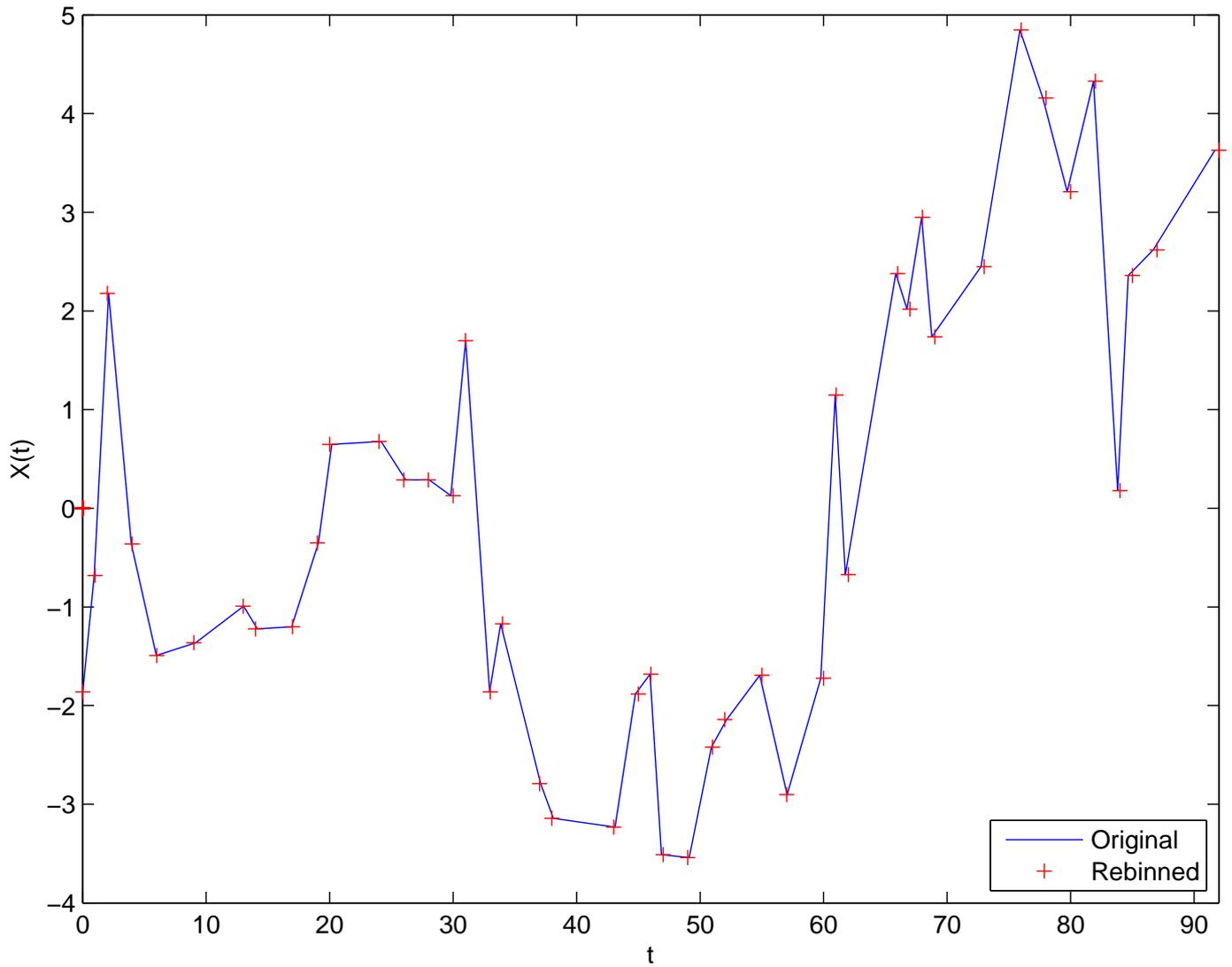}}
        \caption{Experimental (mean-subtracted) time series containing $45$ unevenly-spaced data versus a rebinned version computed on a regular 
        grid of $92$ time instants \citep[data taken from][]{big01}.}
        \label{fig:data1}
\end{figure*}

\begin{figure*}
        \resizebox{\hsize}{!}{\includegraphics{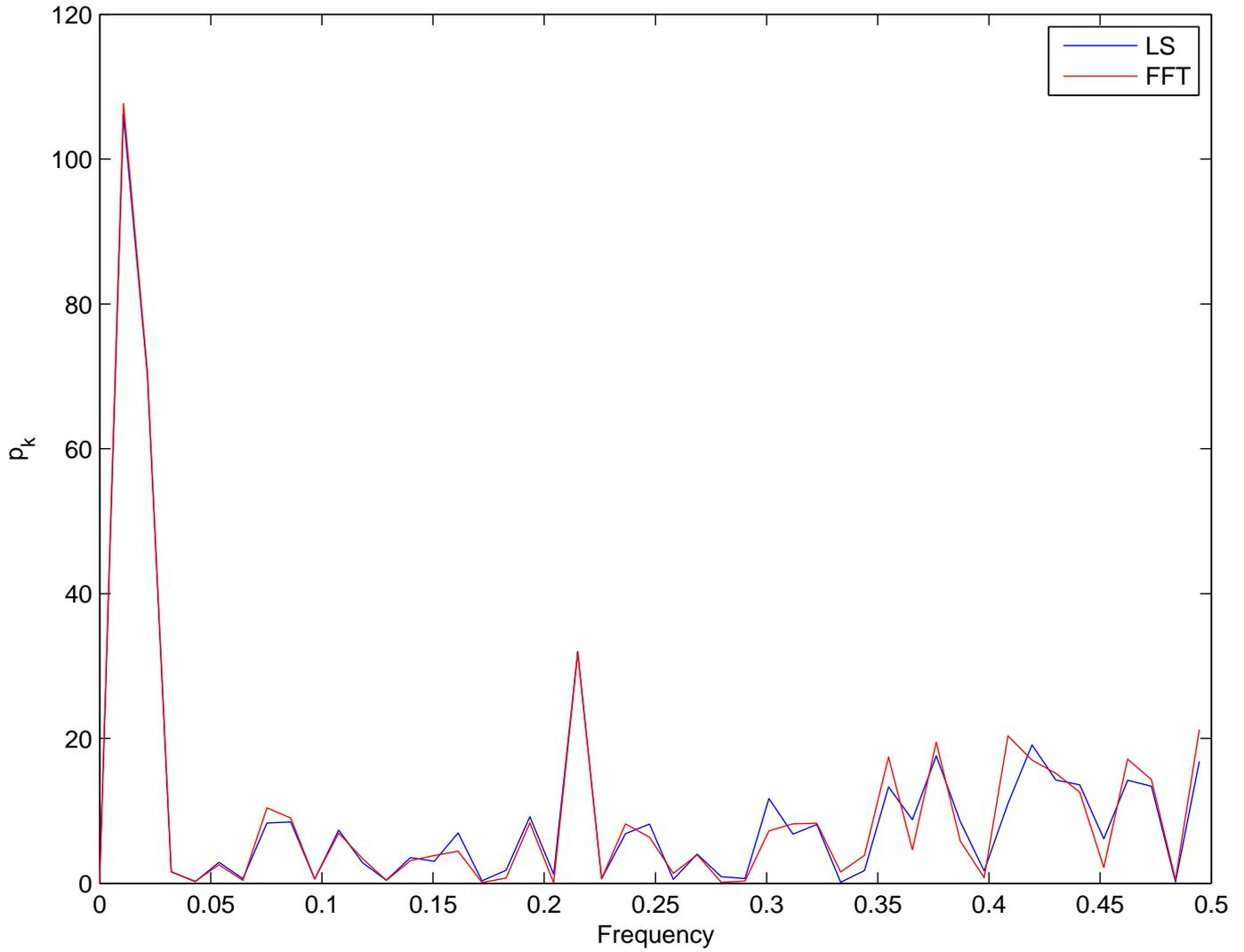}}
        \caption{Periodograms of the time series shown in Fig.~\ref{fig:data1}. Frequency is in given in units of the Nyquist frequency
        corresponding to the shortest sampling time step. The periodogram of the original time series has been obtained
        by means of the {\it Lomb-Scargle} method and that of the rebinned version by means of a classic FFT algorithm.}
        \label{fig:data2}
\end{figure*}

As an example of an unsophisticated method able to produce results
similar to those obtainable with the {\it Lomb-Scargle} periodogram,
we consider the rebinning of the original time series on an
arbitrarily dense regular time grid (in this way a signal with a
regular sampling is obtained but some data are missing) followed by
applying any of the {\it fast Fourier transform} (FFT)
algorithms available nowadays. Figure~\ref{fig:data1} shows an
experimental (mean-subtracted) time series versus its rebinned
version. This time series, which is characterized by rather irregular
sampling, was obtained with the VLA array \citep{big01} and consists
of polarisation position angle measurements at an observing frequency
of 15~GHz for one of the images of the double gravitational lens system
B0218+357. The original sequence contains only $M=45$ data and it is
rebinned on a regular grid of $M_r = 92$ time instants. In spite of
this, as visible in Fig.~\ref{fig:data2}, the corresponding
periodograms, computed on $N=M_r$ equispaced frequencies by means of
the {\it Lomb-Scargle} and the FFT approach, are remarkably similar.
Here, the highest frequency approximately corresponds to the Nyquist
frequency that is related to the shortest sampling time step. The
main conclusion of this example is to point out that, although 
in the previous section we stated that use of the {\it Lomb-Scargle} periodogram 
can be expected to be effective only for time series that contain a small number of data, this is not a sufficient
condition to guarantee that the method is truly useful.

\section{Conclusions}

In this paper we worked out a general formalism, based on the matrix
algebra, that is tailored to analysis of the statistical
properties of the {\it Lomb-Scargle} periodogram independently of the characteristics of
the noise and the sampling. With this formalism it has become possible
to develop a test for the presence of components of interest in a signal in more general
situations than those considered
in the current literature (e.g. when noise is colored and/or
non-stationary). Moreover, we were able to clarify the relationship between the {\it Lomb-Scargle} periodogram
and other techniques (e.g.\ the least-squares fit of sinusoids)
and to fix the conditions under which the use of such method can be expected to be effective.

\clearpage
\clearpage
\clearpage

\end{document}